# General framework for incoherent topological structured light and optical information encoding

Ao Zhou[1,2], Dong Xu[1,2], Yaning Zhou[1,2], Peng Li[3], Pujuan Ma[1,2], Xin Liu[1,2,*], Lin Liu[4], Jianlin Zhao[3], Zhigang Chen[5,*], Yangjian Cai[1,2,*], and Chunhao Liang[1,2,*]

[1]Shandong Provincial Key Laboratory of Light Field Manipulation Physics and Applications & School of Physics and Optoelectronics, Shandong Normal University, Jinan 250358, China.

[2]Collaborative Innovation Center of Light Manipulations and Applications, Shandong Normal University, Jinan 250358, China

[3]Key Laboratory of Light Field Manipulation and Information Acquisition, Ministry of Industry and Information Technology, and Shaanxi Key Laboratory of Optical Information Technology, School of Physical Science and Technology, Northwestern Polytechnical University, Xi' an 710129, China

[4]School of Physical Science and Technology & Collaborative Innovation Center of Suzhou Nano Science and Technology, Soochow University, Suzhou 215006, China

[5]The MOE Key Laboratory of Weak-Light Nonlinear Photonics, TEDA Applied Physics Institute and School of Physics, Nankai University, Tianjin 300457, China

*Corresponding authors: xinliu@sdnu.edu.cn; zgchen@nankai.edu.cn; yangjiancai@sdnu.edu.cn; chunhaoliang@sdnu.edu.cn TEL: +86 15051591677

## Abstract

**Topology provides a powerful language for describing global invariants in physical systems, yet optical topology has been explored predominantly with**

**fully coherent light. Recent studies have shown that incoherent light can host topological structures mediated by coherence singularities; however, a general framework for their construction and control has been lacking. Here, we introduce an incoherent Milnor polynomial, which establishes a theoretical framework for real-space incoherent topological structured light, in which topology and statistical coherence emerge as independent and jointly addressable degrees of freedom. This framework overcomes a fundamental limitation of coherent topological structured light, enabling arbitrary intensity engineering without altering the underlying topological configuration. Experimentally, we realize incoherent Hopf-linked and trefoil-knotted coherence singularities with programmable statistical coherence. We further demonstrate a robust optical information-encoding scheme inspired by Rubik's-cube-like rotations, where statistical coherence determines far-field intensity patterns associated with the cube's initial states, and topological structures govern controlled rotations acting as encryption keys. Our results advance incoherent topological structured light from a physical curiosity to a programmable photonic platform, opening new avenues for optical information encoding, statistical photonics, and coherence-engineered functionalities beyond coherent optical topology.**

## Introduction

Topology provides a general framework for describing structures that remain invariant under continuous deformations, and has profoundly influenced diverse branches of physics, ranging from fluid dynamics and magnetism to biology and quantum field theory.[1-3] In optics, a series of theoretical and experimental breakthroughs over the past two decades have led to the creation of increasingly sophisticated real-space optical topological structures, ranging from the optical vortex knots threaded by the stranded helices[4] to isolated optical vortex knots,[5] optical Hopfion crystals,[6] optical skyrmions,[7,8] optical hopfions,[9] quantum topological photonics,[10] and polarization Möbius strips.[11] These innovations have been propelled by advanced wavefront-engineering techniques, such as holography and metasurfaces,[12-14] enabling the generation of reconfigurable knots and links across length scales spanning from macroscopic to micro- and nanoscale regimes. Moreover, their robustness against moderate aberrations and weak turbulence has highlighted their promise for information encoding, encryption and transfer in realistic environments.[14-18]

Despite these advances, existing studies have been almost exclusively restricted to fully coherent light, where topological features are encoded in well-defined *phase or polarization singularities*. However, the strict coherence requirements underlying such approaches limit their robustness in strongly noisy or turbulent environments, posing a challenge for scalable applications in multiplexed or free-space channels.[19] In contrast, incoherent light, including incoherent topological structured light (ITSL), is expected to exhibit enhanced resilience to turbulence. This is because an incoherent beam can be represented as a statistical superposition of coherent modes, for which turbulence-induced distortions acting on different modes are partially averaged, thereby mitigating the overall structural degradation.[20,21] An incoherent light can be characterized by the cross-spectral density function, $W(\mathbf{r}_1,\mathbf{r}_2)=\tau(\mathbf{r}_1)\tau^*(\mathbf{r}_2)\mu(\mathbf{r}_1,\mathbf{r}_2)$ with $\tau$ and $\mu$ representing the deterministic field amplitude and the complex degree of coherence, respectively.[22] Importantly, the degree of coherence constitutes a unique, independent, and tunable degree of freedom, distinguishing incoherent light from fully coherent light. It couples with other optical degrees of freedom, enabling new insights into light-matter interactions, and emerging applications in information encoding, optical imaging, and optical computing.[23-26] Recently, it has been demonstrated that incoherent light can support topological knots and links encoded in *coherence singularities* hidden within the statistical correlations of speckled fields.[27] Nevertheless, a general theoretical framework for ITSL remains absent.

In this work, we advance incoherent topology from the demonstration of topological phenomena to a general framework by introducing the incoherent Milnor polynomial. We show that its normalized form directly defines the complex degree of coherence, thereby establishing a mathematical and physical bridge among singular optics, real-space topology, and optical coherence. Within this framework, topology and statistical coherence emerge as independent yet jointly programmable degrees of freedom, enabling optical knots/links and statistical coherence distributions to be designed independently within a unified formalism. As a proof of principle, we demonstrate an optical information encoding and decoding protocol based on these two degrees of freedom. Figure 1 provides a conceptual overview of this protocol: the source statistical coherence determines the far-field intensity pattern, which defines the initial Rubik's-cube state, whereas the reconstructed topological structure is associated with a prescribed braid word that serves as the encryption key. This braid word determines a sequence of rotational operations in the cube model, and the decoded information is extracted from the front face of the final cube state. Detailed mappings among the braid word, statistical coherence, far-field intensity pattern, cube state and information recovery are provided in Figs. 2–4 and Supplementary Notes 1, 2, 4 and 5.

This integrated topological–statistical strategy elevates incoherent topology from a passive descriptor into an active design principle, opening new routes towards coherence-engineered photonic functionalities in optical encryption, robust communication and information processing.

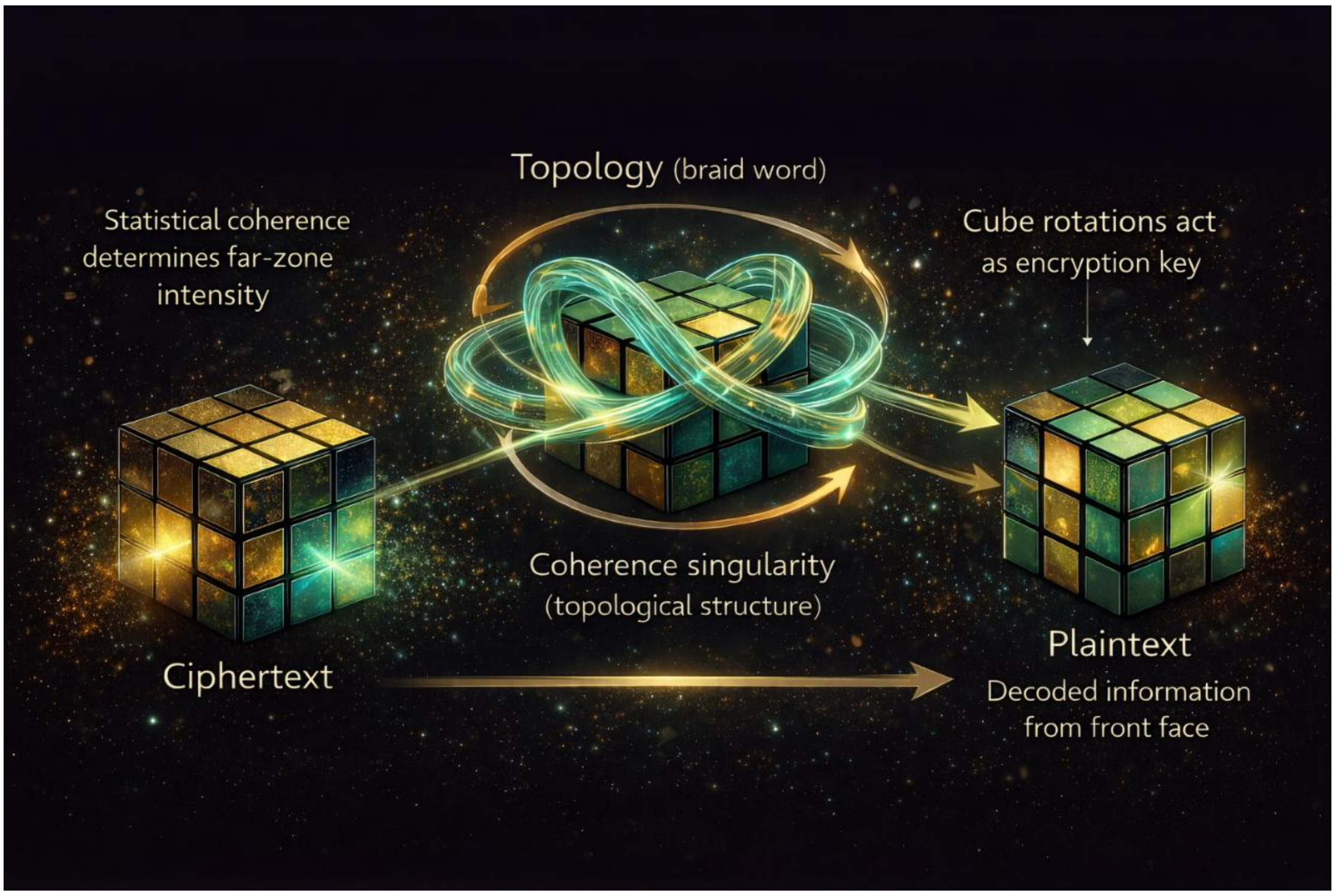

**Fig. 1 | Concept of an optical information architecture based on topology and statistical coherence.** Information decoding is achieved jointly by statistical coherence and real-space topology, where six ITSL beams with distinct source

coherence define the cube faces via their far-field intensity patterns, forming the ciphertext. Topological structures of coherence singularities, described by braid words, act as keys by inducing a sequence of cube rotations. Each encoded state is thus carried by multiple ITSL beams that share the same topological configuration but differ in statistical coherence, enabling robust and adaptable optical information encoding. Information encoding proceeds through the inverse operations.

## Results

### Theoretical results

Unlike real-space coherent topological structured light based on phase and polarization singularities, ITSL encodes topology in coherence singularities, where the complex degree of coherence vanishes. These singularities do not enforce zero intensity, but instead mark regions where coherence disappears while the surrounding field remains bright, revealing a distinct, coherence-driven form of optical topology. To formulate incoherent topology mathematically, we generalize the braid representation method[5] by introducing a stochastic degree of freedom. In the coherent setting, a periodic N-strand braid can be described by the roots of the polynomial in the complex variable *u*, $p_h(u)=\prod_{s=1}^{S}\left[u-Z_s(h)\right]$, where $Z_s(h)=X_s(h)+iY_s(h)$. $X_s$ and $Y_s$ are trigonometrically parameterized by the height *h* (see Fig. S1a). To generalize this framework to incoherent fields, we introduce a random disturbance $\Phi$ and yields $P_h(u)=(p\circ\Phi)(u,h)$, where " $\circ$ " denotes function composition and induces a stochastic deformation of the underlying braid structure (Fig. S1b). To streamline both the mathematical formulation and the physical interpretation, we employ a class of random braid polynomials: $P_h(u) = p_h(u) \cdot \alpha(u,h)$, where $\alpha(u,h)$: $\mathbb{C} \times (0, 2\pi]$ is a random function. Such a random function yields a statistically stationary polynomial, whose statistical properties are characterized by a second-order correlation function:

$$\Omega(u_1, h_1; u_2, h_2) = \langle P_{h_1}(u_1) P^*_{h_2}(u_2) \rangle = p_{h_1}(u_1) p^*_{h_2}(u_2) \eta(u_1, h_1; u_2, h_2), \tag{1}$$

where the angle bracket denotes the ensemble average, and $\eta(u_1, h_1; u_2, h_2) = \langle \alpha(u_1, h_1) \alpha^*(u_2, h_2) \rangle$. Upon replacing $h_n$ with $v_n = \exp(ih_n)$, where $n = 1, 2$, $\Omega(u_1, h_1; u_2, h_2)$ is rewritten as a statistical polynomial $Q(u_1, v_1; u_2, v_2)$. By applying stereographic projection,

$$u = \frac{x^2 + y^2 + z^2 - 1 + 2iz}{x^2 + y^2 + z^2 + 1}, \ v = \frac{2(x + iy)}{x^2 + y^2 + z^2 + 1}, \tag{2}$$

we obtain the incoherent Milnor polynomial $F(\mathbf{R}_1, \mathbf{R}_2) = Q(u_1, v_1; u_2, v_2)$ where $\mathbf{R} \equiv (x, y, z)$, thereby generating real-space knotted or linked coherence singularities (Fig. S1c). When z=0, the incoherent Milnor polynomial diverges as $x, y \to \infty$. To physically construct the ITSL, we introduce a Gaussian truncation $\exp(-\mathbf{r}^2/2\omega^2)$ with a beam width $\omega$ and an overhomogenization term $(\mathbf{r}^2 + 1)^m$ under different coefficient *m*.[28] At the source plane z = 0, the well-defined cross-spectral density function for ITSL, expressed as:

$$W(\mathbf{r}_1, \mathbf{r}_2) = \exp\left(-\frac{\mathbf{r}_1^2 + \mathbf{r}_2^2}{2\omega^2}\right) (\mathbf{r}_1^2 + 1)^m (\mathbf{r}_2^2 + 1)^m F(\mathbf{r}_1, \mathbf{r}_2), \tag{3}$$

where $\mathbf{r} \equiv (x, y)$. The Gaussian truncation ensures finite beam energy, whereas the overhomogenization term improves the physical realizability and propagation stability of the target vortex knots or links while preserving their prescribed topology.

Based on the coherence theory,[22] the deterministic field amplitude $\tau(\mathbf{r})$ and complex degree of coherence $\mu(\mathbf{r}_1,\mathbf{r}_2)$ of the ITSL are given, respectively, as follows

$$\tau(\mathbf{r}) = \exp\left(-\frac{\mathbf{r}^2}{2\omega^2}\right)\left(\mathbf{r}^2+1\right)^m \sqrt{F(\mathbf{r},\mathbf{r})}, \tag{4}$$

and

$$\mu(\mathbf{r}_1,\mathbf{r}_2) = \gamma(\mathbf{r}_1,\mathbf{r}_2) = \frac{F(\mathbf{r}_1,\mathbf{r}_2)}{\sqrt{F(\mathbf{r}_1,\mathbf{r}_1)}\sqrt{F(\mathbf{r}_2,\mathbf{r}_2)}}, \tag{5}$$

where $\gamma$ is the normalized form of the incoherent Milnor polynomial. It is required to be a Hermitian and non-negative-definite kernel,[22] thereby ensuring that the constructed field corresponds to a physically realizable partially coherent beam. Equation (5) provides the central physical insight of this work: the normalized incoherent Milnor polynomial explicitly determines the complex degree of coherence, forming the mathematical and physical foundation of ITSL. The polynomial $\gamma(\mathbf{r}_1,\mathbf{r}_2)$ or the function $\mu(\mathbf{r}_1,\mathbf{r}_2)$ simultaneously encodes topological configuration and statistical coherence, while treating them as independent degrees of freedom: the braid function $p_h(u)$ determines the topological skeleton of coherence-singularity structures, whereas the random function $\alpha(u,h)$ or statistical kernel $\eta(u_1,h_1;u_2,h_2)$ governs the statistical coherence. As a result, topological structures and coherence distributions can be designed independently within a unified formalism. The mathematical derivation and controllability analysis are provided in Supplementary Notes 1 and 2. This independent programmability constitutes the key distinguishing feature of the proposed framework. By combining the parametrized braid construction with a general statistical-coherence description, the incoherent Milnor polynomial provides a systematic route for generating a broad family of optical knots and links with prescribed coherence properties. Unlike previous studies that demonstrated specific ITSL [27], the present work establishes a general framework for constructing and programming incoherent topology. Furthermore, coherent topological structured light[5] and recently reported Gaussian-correlated ITSL [27] emerge naturally as special cases, obtained by choosing the random function $\alpha(u,h)$ to be deterministic or to generate a Gaussian coherence kernel, respectively. These results elevate incoherent topology from the demonstration of individual phenomena to a constructible and programmable framework for ITSL and its applications in information processing.

The schematic diagram of the ITSL during propagation in free space is given in Fig. 2a. The topological structure is embedded into the complex degree of coherence confined near the source, while the far-field intensity distribution can be flexibly tailored through the source statistical coherence (see Supplementary Note 4), different from the source intensity profile. This breaks the longstanding limitation in real-space topology whereby optical intensity patterns are constrained to fixed configurations dictated by topology. To demonstrate the effect, we adopt two distinct incoherent light beams as detailed in the Supplementary Information, and use the Monte Carlo method for ITSL numerical simulations. We adopt 1000 realizations and they are sufficient to obtain stable distributions of the complex degree of coherence and optical knots/links. The modulus and phase of the degree of coherence are displayed in Fig. 2b2 and 2c2.

Coherence singularities in the phase coincide with vanishing modulus ( $\mu = 0$ ), forming dark regions. By locating these singularities across successive transverse planes along the propagation path and linking them sequentially, the targeted trefoil knot and Hopf link are reconstructed (Fig. 2b1 and 2c1). We employ the Artin braid group B2, and their braid words are respectively given by $\sigma_1^{+3}$ and $\sigma_1^{+2}$ . After far-field propagation, the source statistical coherence further gives rise to the desired intensity distribution (Fig. 2b3 and 2c3). They well demonstrate that the statistical coherence and topological configuration constitute two independent degrees of freedom, which can be manipulated separately to realize tailored structured light.

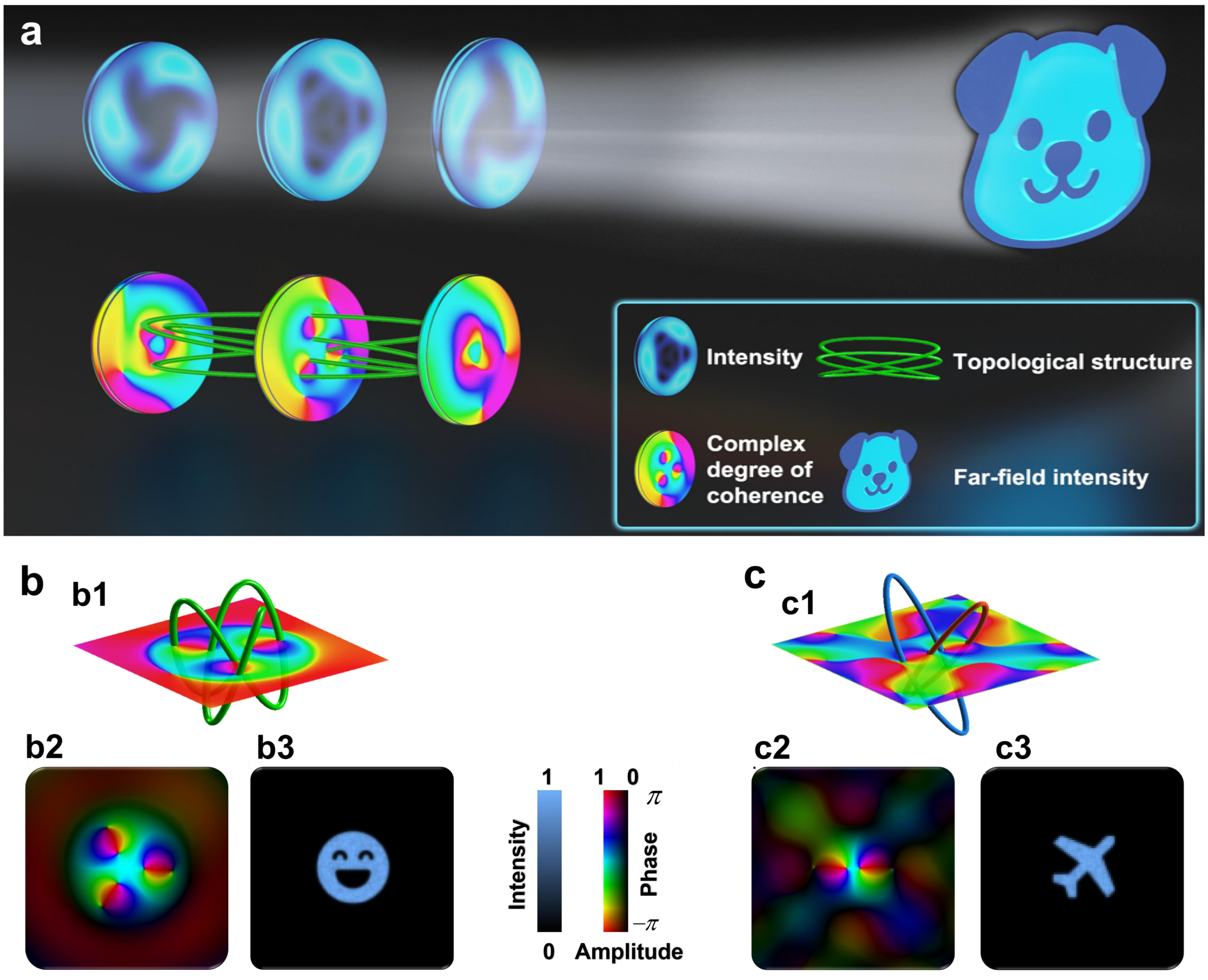

**Fig. 2 | Concept and theoretical framework of ITSL. a**, Schematic of free-space propagation of ITSL, where topology is embedded in the complex degree of

coherence (bottom), while source statistical coherence independently controls the far-field intensity distribution. **b,c**, Reconstructed modulus and phase of the complex degree of coherence at the source plane, showing that coherence singularities coincide with vanishing modulus and form distinct real-space topological configurations, such as a trefoil knot **b1** and a Hopf link **c1**. By tuning the source statistical coherence, identical topological configurations can support different far-field intensity distributions.

We further exploit these two independent degrees of freedom to develop a high-capacity information encoding strategy. As illustrated in Fig. 3, the scheme is conceptually inspired by the rotations of a Rubik's cube. In Fig. 3a, the three-dimensional cube consists of six faces, each composed of 2×2 blocks in two colors (pink and sky blue). Four independent rotation trajectories are defined and denoted as $\sigma_1$, $\sigma_2$, $\sigma_3$, and $\sigma_4$, each with two directions: a superscript "+" for a rightward rotation of the upper and lower rows or an upward rotation of the left and right columns, while the "−" represents the opposite rotations. The cube rotations can be mathematically described by braid words of the topology. For example, the braid word $\sigma_1^-$, i.e., rotating the first row clockwise once. Through careful design of the initial cube states and selection of appropriate topological structures, we successfully realize all 16 distinct color states from the front 2×2 blocks (see Supplementary Note 5). With two colors mapped to binary digits (pink → 0, sky blue → 1), the 16 encoded states are achieved ("0000"–"1111"). In our physical implementation, we replace the sub-squares in two colors with optical patterns:

quarter circles represent "sky blue" sub-square while L-shape structures represent "pink" sub-square. We express 16 encoded states by the structured intensity patterns, with several examples shown in the lower-right corner of Fig. 3a.

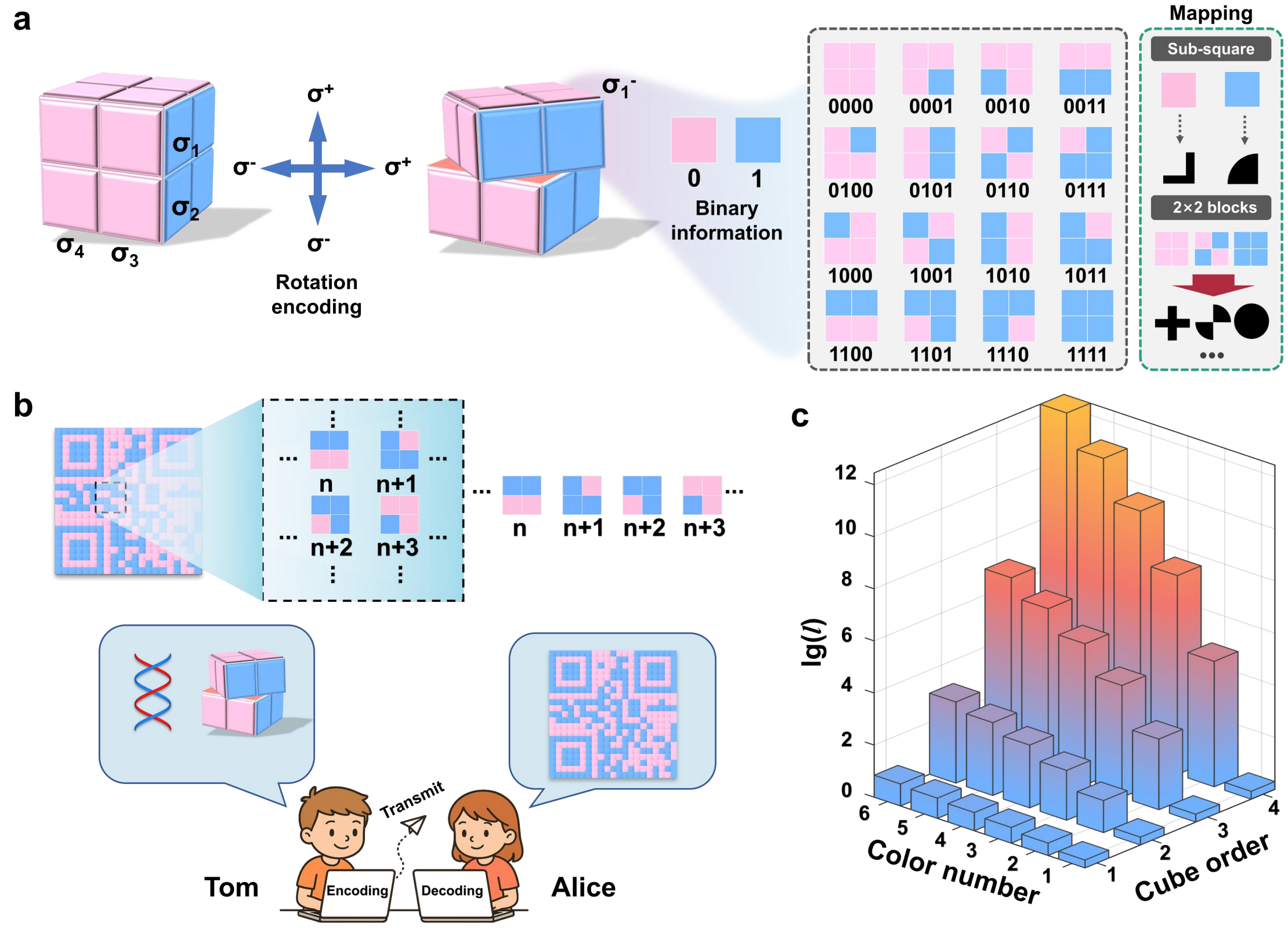

**Fig. 3 | High-capacity optical information encoding using ITSL. a**, Conceptual encoding architecture inspired by Rubik's-cube rotations, where each face consists of 2×2 blocks rendered in two colors (pink and sky blue). Discrete rotations of rows or columns along four prescribed trajectories ( $\sigma_1 - \sigma_4$ ) generates 16 distinct color configurations on the front face, corresponding to hexadecimal information states. For physical implementation, the two-color sub-squares are mapped to different structured patterns, which are used to represent and characterize all encoded states. **b**, Schematic of QR-code encryption and decryption implemented with ITSL, illustrating information encoding and recovery through controlled topological operations. **c**,

Scaling of information capacity with cube order and color number, showing exponential growth of the encoding space, which enables high-capacity optical information encoding and transmission.

In the encoding and decoding protocol, the far-field intensity pattern, governed by the source's statistical coherence, specifies the initial cube state and serves as the ciphertext. The braid word functions as the encryption key, determining the final cube state through a sequence of rotational operations. Since different braid words can represent the same closed knot or link under Markov equivalence, a reconstructed topology does not uniquely determine a braid word. In our protocol, this ambiguity is avoided by predefining a finite topological alphabet, where each selected knot or link is assigned a representative braid word as its key. For example, although a trefoil knot can be represented by different closed braids, such as the closure of $\sigma_1^{+3} \in B_2$ or the closure of $\left(\sigma_1\sigma_2\right)^2 \in B_3$, our scheme assigns the trefoil knot to the canonical braid word $\sigma_1^{+3}$. The decoded information is extracted from the front face of the final cube state, which is obtained through a sequence of rotational operations defined by braid word. Within this framework, the overall information of the cube is determined by its six faces, which are represented by six incoherent light beams with distinct sources' statistical coherence. The cube rotations are encoded by the corresponding topological structures. Consequently, each encoded state is jointly carried by six ITSL beams that share the same topological configuration but differ in their statistical coherence. The above procedure represents the information extraction process, and the inverse operation can be performed for information encoding.

We illustrate the protocol using an example where Tom intends to send an encrypted QR code to Alice over a long distance (Fig. 3b). The QR code is first divided into a series of 2×2 blocks arranged in sequential order. Each block is then mapped onto the color distribution of the Rubik's cube front face. We determine the topological configuration and then reconstruct the initial cube state via reverse rotational operations. Such encoded Rubik's cube is represented by six incoherent light beams that share the same topological structure but possess different statistical coherences. At the receiver's end, Alice directly measures the far-field intensity to obtain the cube's initial state. The topological configuration (braid word) can be reconstructed using diffraction or interference methods.[29-31] By combining the braid word with the initial cube state, Alice can recover the blocks in sequence, in turn, reconstruct the encrypted QR code.

We further point out that the information capacity can be exponentially enhanced by increasing the order of the Rubik's cube and the number of available colors. As shown in Fig. 3c, for a 3×3 cube with two colors, the front face consists of 3×3 = 9 sub-squares, yielding 2^9 = 512 possible encoded states $l$. Extending to a 4×4 cube with six colors, the number of possible states grows dramatically to $l$ = 6^16 ≈ 2.8×10^12. This result demonstrates that, by simply scaling the cube order and color set, the information capacity of this dual-encoding approach can expand exponentially, offering enormous potential for high-capacity information encryption and transmission.

**Experimental demonstration**

In the experiment, we first generated a set of ITSL realizations using the Monte Carlo method and, via complex-amplitude modulation, encoded them into a holographic video. This video was then uploaded to a liquid-crystal spatial light modulator (SLM), where the target ITSL was produced through the incoherent superposition of the +1 or −1 diffracted orders. To enable field reconstruction, we further introduced a reference beam using off-axis holography, allowing it to interfere with individual ITSL realizations. From the resulting interference patterns, the complex field of each realization was retrieved. All reconstructed realizations were then synthesized to obtain the source cross-spectral density function, and the global experimental topological structure was reconstructed using the angular spectrum method[29] combined with coherence singularity tracking (see Supplementary Note 8).

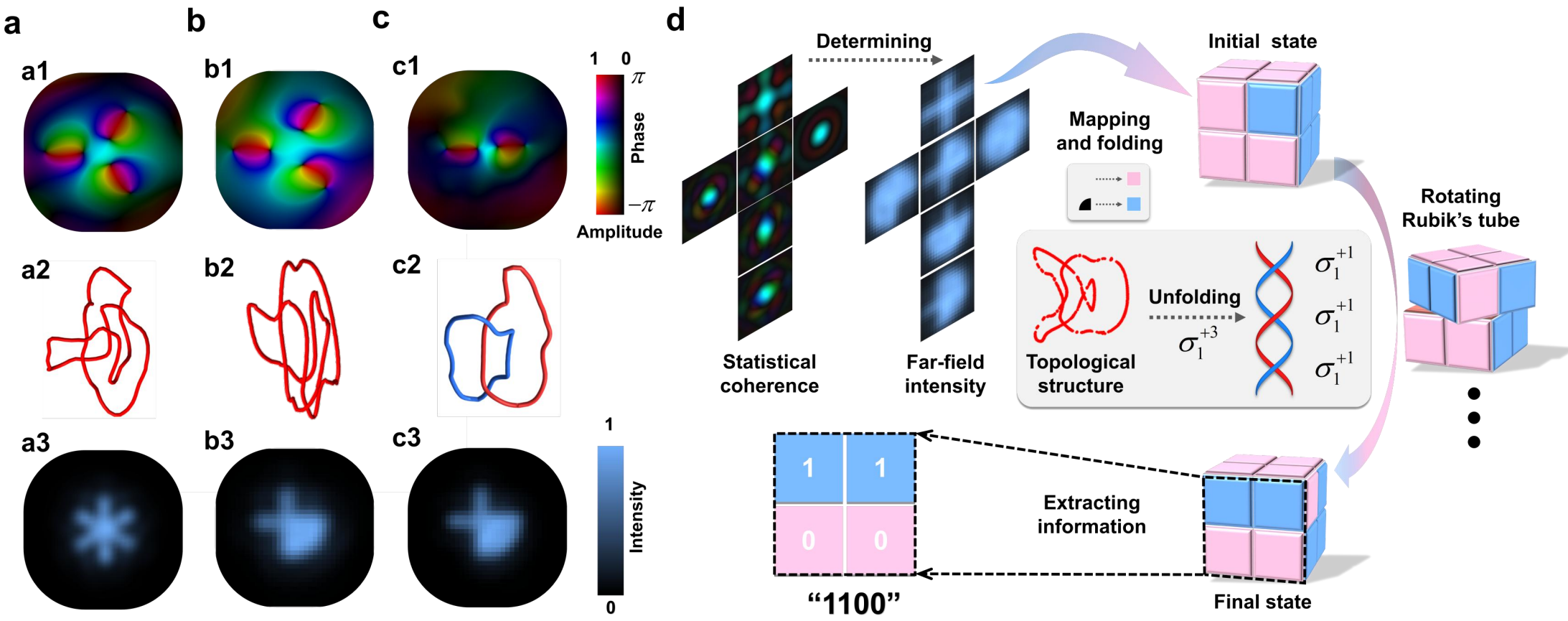

**Fig. 4 | Experimental generation, information encryption, and recovery of ITSL.** **a1,b1,c1**, Reconstructed modulus and phase of the complex degree of coherence at the source plane, revealing distinct and robust coherence singularities. **a2,b2,c2**, Reconstructed three-dimensional coherence singularities, exposing real-space topological configurations, such as trefoil knots **a2,b2** and a Hopf link **c2**. **a3,b3,c3**,

Corresponding far-field intensity distributions under different source statistical coherence, demonstrating independent and tunable optical responses without altering the underlying topology. **d**, Experimental demonstration of information encryption and recovery using ITSL, enabled by the joint control of statistical coherence and topological structure.

Figures 4a1-4c1 present the experimentally reconstructed modulus and phase of the complex degree of coherence at the source plane. The corresponding theoretical models for these incoherent light beams are provided in the Supplementary Information. Despite the coherence modulation that induces local phase distortions, coherence singularities remain distinct and robust. We further experimentally revealed topological configurations, including trefoil knots (Fig. 4a2 and 4b2) and a Hopf link (Fig. 4c2). Their braid words in Artin braid group B2 are given by $\sigma_1^{+3}$ and $\sigma_1^{+2}$, respectively. Through devising the source statistical coherence, the far-field intensity patterns could be customized on demand (Figs. 4a3-4c3). It is worth noting that in Figs. 4a and 4b, we use the same topological structure but different source statistical coherence, whereas in Figs. 4b and 4c, we use different topological structures under the same source statistical coherence. These two complementary comparisons experimentally demonstrate, in a more direct manner, that topology and statistical coherence constitute two independently and programmably controllable degrees of freedom. It well illustrates the controllable nature of ITSL by tuning the normalized incoherent Milnor polynomial or complex degree of coherence function.

Additionally, we experimentally conducted an example of information encryption and transmission. We utilized six incoherent light beams with different source statistical coherence (Fig. 4d), sharing the same topological structure (trefoil knots) for information encoding. The statistical coherence of these light beams determines the far-field intensity distribution, which in turn dictates the initial state of the Rubik's cube. All experimental details are provided in Fig. S10. The trefoil knot, described by the braid word $\sigma_1^{+3}$ in Artin braid group B2, is used to rotate the first row of the cube counterclockwise three times and determine its final state, from which information "1100" is extracted through the front face. Due to the robustness of the topological structure and its inherent difficulty in being accessed through coherent phase, this information encoding technique naturally possesses high fidelity and high security.

Although the preceding theoretical analysis suggests that the encoding capacity can, in principle, be expanded without a strict upper bound by increasing the cube order, the practically usable capacity is inevitably constrained by experimental and computational limitations. In our encoding protocol, the independent rotation trajectories of the cube are represented by braid generators $\sigma_i$ . Therefore, since the cube order is equal to half the number of independent rotation trajectories, its maximum achievable value is ultimately constrained by the number of braid strands. Increasing the cube order consequently requires optical knots or links with higher topological complexity. More complex topological configurations produce denser and more closely spaced coherence singularity trajectories, making their reconstruction and classification more sensitive to experimental noise, detector resolution, SLM pixel size, and external perturbations. Decoding requires retrieving the complex degree of coherence, tracking the singularity lines, identifying the knot or link type and matching it to a predefined braid word dictionary. Therefore, the protocol exhibits a capacity–complexity trade-off: a larger topological alphabet increases the nominal encoding capacity, but also raises the experimental and computational cost of reliable decoding.

## Discussion

This work establishes ITSL as a new frontier at the intersection of topology, singular optics, and statistical optics. In contrast to coherent fields, where topology is intrinsically locked to deterministic phase or polarization structures, ITSL encodes topology within coherence singularities, allowing the statistical nature of light to play a constructive and functional role. The normalized incoherent Milnor framework provides a unified and general description that separates and independently controls two fundamental degrees of freedom: real-space topological configuration and statistical coherence.

Experimentally, we verified these concepts by generating trefoil-knotted and Hopf-linked coherence singularities whose topology remains robust under systematic variation of statistical coherence. This robustness arises because the topology is defined by the zero lines of the ensemble-averaged complex degree of coherence, rather than by the fluctuating phase of a single realization. Accordingly, the stable coherence region, characterized by the transverse coherence width, must encompass the spatial extent of the topological structures. The demonstrated independence between topology and coherence opens new opportunities for noise-resilient photonic systems and coherence-controlled optical functionalities. For example, dynamic modulation of statistical coherence could enable adaptive imaging systems,[32] turbulence-immune communication channels[33] and optical information encoding,[34,35] leveraging the inherent statistical stability of incoherent light.

Beyond classical optics, this framework provides a conceptual foundation for exploring topological phenomena in quantum coherence,[36] offering new insights into the interplay between statistical fluctuations and topological invariants. The concept of joint coherence–topology encoding may also advance photonic computing and secure multiplexing, complementing recent developments in incoherent neural and computational photonics.[37]

Recent progress in optical vortices[38,39] and structured light[40] has provided a powerful platform for advancing ITSL toward more complex coherence singularity networks, higher-order knot/link structures and multidimensional information channels in ITSL. Further, ITSL could also be integrated with metasurfaces[41] and spatiotemporal modulators[42-44] to achieve miniaturized and ultrafast programmable topological architectures, or extend to other wave systems, including acoustics and matter waves. More broadly, the integration of structured light, statistical coherence, and topology establishes a versatile paradigm for statistical–topological wave engineering, in which coherence itself becomes a programmable degree of freedom for encoding, protecting, and manipulating information..

## Materials and Methods

**Generation and characterization of ITSL**: ITSL beams were generated and characterized using the experimental setup shown in Fig. S9. A linearly polarized laser beam was split into two paths by a beam splitter. The reflected path was modulated by a spatial light modulator (SLM) and directed through a 4f filtering system composed of two lenses and an aperture, which selectively transmitted either the +1 or -1 diffracted order. The SLM was loaded with a designed holographic video. The filtered output served as the dynamic realizations, which were subsequently focused by a lens and incoherently superposed to obtain the far-field intensity distribution.

An off-axis holography technique was then employed, in which a reference beam from the splitter was redirected by two mirrors to interfere with the target realizations. The resulting interference fringes enabled retrieval of individual realizations and subsequent reconstruction of the source cross-spectral density function. The complex degree of coherence of the ITSL beams at different propagation distances was obtained through numerical reconstruction. The target topological structures were finally reconstructed using a coherence singularity tracking method. All experimental details required to reproduce the results are provided in the Supplementary Notes 6-11.

**Data availability**

All data are displayed in the main text and Supplementary Information. The data that support the findings of this study are available from the corresponding author upon reasonable request.

**Acknowledgements**

This work was supported by the National Key Research and Development Program of China (2022YFA1404800), the National Natural Science Foundation of China (12534014, 12374311, 12574328, 12192254, W2441005, W2541003,12134006 and 12547149), the Taishan Scholars Program of Shandong Province (tsqn202312163) and the Natural Science Foundation of Shandong Province (ZR2025ZD21, ZR2024QA012 and ZR2023YQ006)

**Author contributions**

C.L. conceived the research. A.Z., D.X., Y.Z. and X.L. verified the theoretical framework and performed the simulations and experiments. P.L. P.M. L.L. completed the data analysis and visualizations. A.Z., X.L., J.Z., Z.C., Y.C. and C.L. contributed to drafting and finalizing the manuscript. X.L., Z.C., Y.C. and C.L. co-supervised the project. All authors participated in the discussion and review of the manuscript.

**Conflict of Interest**

The authors declare no competing interests

Supplementary information accompanies the manuscript on the Light: Science & Applications website (http://www.nature.com/lsa)